\documentclass[twocolumn, times]{aastex631}

\accepted{\today}

\submitjournal{AJ}

\begin{document}

\title{Residual Entropy as a Diagnostic and Stopping Metric for CLEAN}

\correspondingauthor{D. C. Homan}
\email{homand@denison.edu}

\author[0000-0002-4431-0890]{D. C. Homan}
\affiliation{Department of Physics and Astronomy, Denison University, Granville, OH 43023, USA}

\author[0009-0009-5367-5603]{J. S. Roth}
\affiliation{Department of Physics and Astronomy, Denison University, Granville, OH 43023, USA}

\author[0000-0002-9702-2307]{A. B. Pushkarev}
\affiliation{Crimean Astrophysical Observatory, 298409 Nauchny, Crimea, Russia}

\begin{abstract}
We propose the use of entropy, measured from the spatial and flux distribution of pixels in the residual image, as a potential diagnostic and stopping metric for the CLEAN algorithm.  Despite its broad success as the standard deconvolution approach in radio interferometry, finding the optimum stopping point for the iterative CLEAN algorithm is still a challenge.  We show that the entropy of the residual image, measured during the final stages of CLEAN, can be computed without prior knowledge of the source structure or expected noise levels, and that finding the point of maximum entropy as a measure of randomness in the residual image serves as a robust stopping criterion.  We also find that, when compared to the expected thermal noise in the image, the maximum entropy of the residuals is a useful diagnostic that can reveal the presence of data editing, calibration, or deconvolution issues that may limit the fidelity of the final CLEAN map.

\end{abstract}



\section{Introduction} \label{s:intro}

\begin{figure*}[ht]
\centering
 \includegraphics[width=0.93\textwidth]{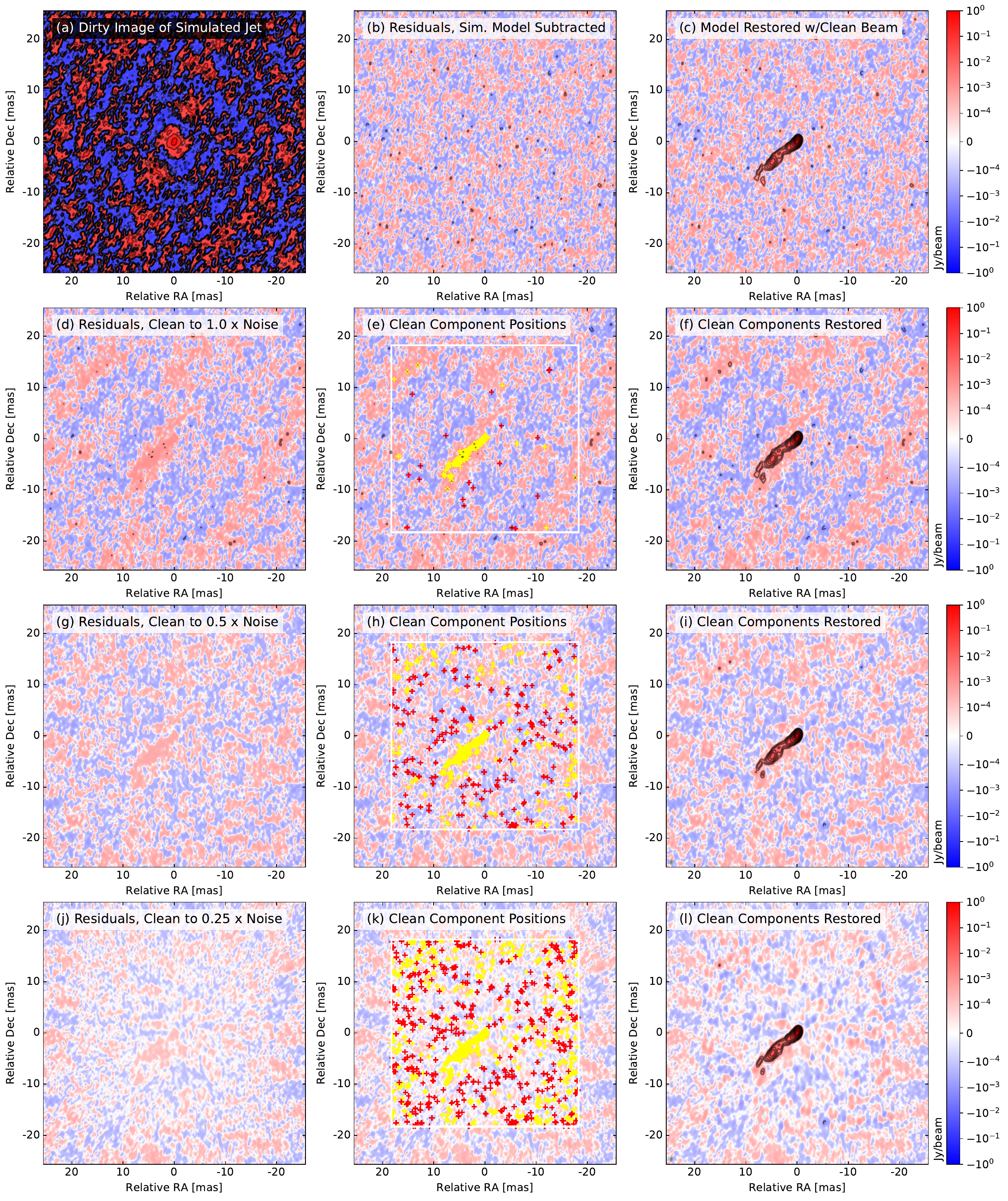}
\figcaption{\label{f:fig1}
  An example of CLEAN carried to different stopping points on simulated data of the 
  parsec-scale jet of NRAO 140.  The first row shows the dirty map (a), residuals if the known model components are removed from the data (b), and restored map using the known model components (c); thus panels (b) and (c) show what we would obtain with a `perfect' deconvolution of the dirty map.  Rows two, three and four show the results of CLEAN after reaching $1.0\times$, $0.5\times$, and $0.25\times$ the known thermal noise in the 
  simulation.  The left and center plots show the residuals with and without CLEAN components
  superimposed (red = negative, yellow = positive).  A single white line indicates the allowed
  cleaning window. The right hand plots show contours with the restored CLEAN maps.  All contours start at $3.0\times$ the known thermal noise of $0.4$ mJy/beam, and increase in 
  factors of 2.0, except for panel (a) which increases in factors of 4.0.
}
\end{figure*}

The CLEAN algorithm by \citet{1974A&AS...15..417H}, along with its
enhancements \citep[e.g.][]{1980A&A....89..377C,1984A&A...137..159S,1984AJ.....89.1076S,2008ISTSP...2..793C}, 
has become the standard deconvolution
approach used in radio interferometry for reconstructing the sky 
brightness distribution from a {\it dirty map} which, in the absence of
noise and calibration errors, consists of the true brightness 
distribution convolved with the {\it dirty beam}. The dirty beam
is the response of the interferometer to a point source, and in a 
sparse array will include significant sidelobes that obscure the 
source structure, see Figure \ref{f:fig1}a.  CLEAN is an iterative
algorithm that reconstructs the sky brightness distribution as
a collection of point sources\footnote{Other well defined
structures with a predictable dirty beam pattern can be substituted,
\citep[e.g.][]{2008ISTSP...2..793C}.} convolved with a {\it clean beam}, usually taken to be
 a two-dimensional Gaussian that matches the central portion of the
 dirty beam.  CLEAN finds these point source {\it clean components} by
 operating in an iterative loop that (a) identifies the 
 peak, $I_{peak}$, in the dirty map, (b) subtracts a point source from that 
 location of amplitude $g\times I_{peak}$ convolved with the dirty beam 
 from the current dirty map, where $g\leq1$ is the loop gain and is
 typically in the range of $0.05$ to $0.1$, (c) records the location
 and amplitude of that clean component subtracted, and (d) returns to step 
 (a) using the current residual dirty map.  The algorithm terminates
 when it reaches a stopping criterion, which we will discuss in more 
 detail below, and the subtracted clean components, convolved with 
 the Gaussian clean beam, are restored to the residual map to make
 the final {\it clean map}.

Unfortunately, it is difficult to determine a well defined stopping criterion
for CLEAN. \citet{1974A&AS...15..417H} recommends stopping CLEAN iterations when the
current value of $I_{peak}$ in the residual map is ``no longer significant
in view of the general noise level of the map''.  In their summary of the 
CLEAN algorithm, \citet{2017isra.book.....T} suggest comparing the residual
$I_{peak}$ to the current Root-Mean-Square (RMS) of the residual map, identifying 
when the RMS level of the residual map fails to continue to decrease, or stopping when
an unacceptable number of negative clean components are identified (presumably 
in Stokes-$I$ which should be positive).

Figure \ref{f:fig1} illustrates three possible stopping 
points for the CLEAN algorithm using simulated data of observations 
of NRAO 140 at 15 GHz with the National Radio Astronomy Observatory's Very Long 
Baseline Array (VLBA).  The simulated data themselves are described in appendix 
B of \citet{2012AJ....144..105H}, and because it is a simulation, we know 
the thermal noise, added in the UV-plane, is equivalent to $0.4$ mJy/beam in a 
naturally weighted image and that there are no other antenna 
gain or baseline based errors in the data.  The first row of Figure \ref{f:fig1} 
shows the original dirty map (left), the residual map if the correct source model is 
subtracted (center), and the final map if the correct source model is restored with 
the clean beam (right).  The remaining rows show the result of the CLEAN 
algorithm stopped when the RMS level of the residuals reach $1.0\times$, 
$0.5\times$, and $0.25\times$ the expected thermal noise.  Notice that in each 
case some remaining source structure can still be seen in the residual map, and it is 
unclear from casual inspection which stopping point would produce the best 
agreement with the known source model. Also notice that as the CLEAN proceeds, 
the apparent noise in the restored clean map decreases, both within the large CLEAN 
window and in the area around that window.  Determining the best stopping 
point therefore impacts not only the correctness of the source model 
but also the estimate of the map noise one would deduce from the final image.

A stopping criterion based on the RMS level of the residual map is attractive
as this quantity is easy to calculate, and one can then compare it to the 
expected thermal noise in the data.  For large monitoring programs, having
an automated stopping criterion is essential, not just for uniformity over time,
but also to allow re-imaging of the entire database without direct human decision-making
on when to stop thousands of CLEAN runs.
Indeed, this is the current approach taken by the MOJAVE program\footnote{https://www.cv.nrao.edu/MOJAVE/} 
\citep[e.g., ][]{2005AJ....130.1389L, 2018ApJS..234...12L} which in most cases stops the CLEAN of an image when the RMS of the residuals 
reaches the thermal noise estimate for those observations. Thermal noise estimates 
require accurate data weights which can be estimated from the scatter within a scan; however, 
these do not include more slowly changing sources of uncertainty like antenna gain errors
from self-calibration, baseline based errors, residual D-terms in polarization data, and 
even errors during the deconvolution process itself.  All of these additional sources of 
uncertainty contribute to the error budget and may limit the ability of CLEAN to extract 
meaningful source structure from the residuals.

In this paper we propose a new approach to monitoring the end of the CLEAN process
by measuring the entropy of the residual map as a whole in the final iterations of
CLEAN.  We emphasize that this is not a variation on the 
Maximum Entropy Method (MEM) of deconvolution \citep[e.g.][]{1978Natur.272..686G, 1985A&A...143...77C}.  Indeed, the entropy value we calculate 
is not used in the CLEAN decision making of what components to remove. Rather we use 
entropy of the residuals as a measure of the {\em randomness} of the residual map as
we approach the point where all significant source structure has been removed from 
the dirty map. Maximizing entropy will favor a Gaussian flux distribution
for the residuals given a fixed mean and variance, and while we do not attempt a theoretical
demonstration that this is the expected distribution for CLEAN residuals, our tests show that the maximum entropy point makes a reasonable stopping criterion in practice.  To this end, we use the information theoretic formulation of entropy 
proposed by \citet{shannon} and equivalent to that used in Statistical 
Mechanics: $S=-\sum p_i \ln p_i$ which \citet{2001A&A...368..730S} identified as
having the best overall properties of the entropy functions typically used in astronomical
data analysis. In section \ref{s:methods} we describe how
we calculate entropy from the residual image, and we detail a set of CLEAN tests on 
simulated VLBA data to determine how the entropy of the residuals changes 
near the end of the CLEAN.  In section \ref{s:results} we discuss our 
results which show that the entropy of the residual map climbs to a maximum value 
during the final stages of CLEAN, and show that this maximum entropy point is a 
sensible stopping criterion that does not require a-priori knowledge of the 
thermal noise or other sources of randomizing uncertainty in the observation.
Finally, in section \ref{s:summary}, we summarize our conclusions.

\section{Methods} \label{s:methods}

We calculate the entropy, $S$, of the residual map by computing the probability that pixels 
fall into one of $N_s$ spatial bins and $N_f$ flux bins.  Before binning in flux,
we divide the residual map by its current RMS taken over the whole field.  In this
way our entropy calculation is independent of the current RMS value of the map and
only depends on the distribution of pixels relative to that RMS level.  Flux bins are
chosen such that they divide each RMS unit interval, e.g. $\ldots-2$ to $-1$, $-1$ to $0$, $0$ to $+1$, $+1$ to $+2\ldots$, into ten equally sized bins each.  Spatial binning is described in detail below. 
Probabilities are computed by simply counting all of the pixels that fall 
into a specific bin and dividing
by the total number of pixels in the map, and the entropy is calculated with the 
following summation.

\begin{equation}
\label{e:Entropy}
S = -\sum_{j=1}^{N_s}\sum_{k=1}^{N_f} p_{j,k}\ln p_{j,k} 
\end{equation}

\begin{figure*}[ht]
\centering
 \includegraphics[width=0.93\textwidth]{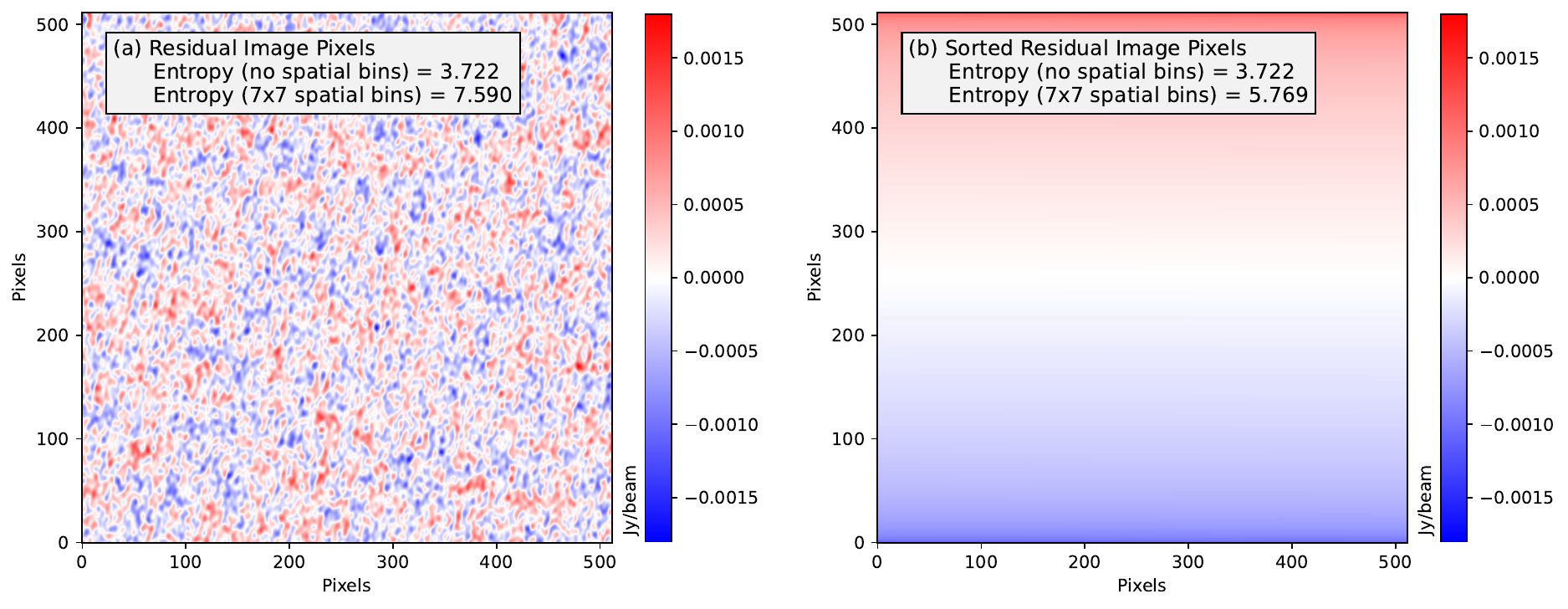}
\figcaption{\label{f:fig2}
An illustration of the effects of spatial binning while computing entropy of a residual map.
Panel (a) is the same residual image shown in figure \ref{f:fig1}b, while panel (b) takes those
pixels and simply sorts them from largest to smallest.  An entropy calculation that does not
include spatial bins cannot distinguish the two panels, but including spatial information
reveals panel (a) to have significantly higher entropy than panel (b).
}
\end{figure*}

The use of spatial bins allows our entropy calculation to naturally detect areas
of the image that contain different flux distributions.  Figure \ref{f:fig2} illustrates
this property of our entropy calculation by showing the same set of pixel fluxes in two different
spatial arrangements.  Figure \ref{f:fig2}a is simply the residual fluxes from panel 
\ref{f:fig1}b, and figure \ref{f:fig2}b sorts those same pixels into order from
largest to smallest. As illustrated in the figure, an entropy calculation without spatial 
binning cannot distinguish between the two panels, while the entropy calculation that uses 
a $7\times7$ spatial grid finds a significantly larger entropy for the randomly distributed
pixels (panel a) than the sorted pixels (panel b).

The size of the spatial bins are picked to encompass a large number of independent beams 
while still being small enough to distinguish residual source structure from other 
locations in the map.  For all the results discussed in this paper, we chose a $7\times7$ spatial
binning grid in our 512 x 512 pixel images, giving approximately one hundred beam areas per spatial bin\footnote{The pixels are correlated by convolution with the telescope beam, so we felt it was
important to pick a spatial binning size that had many independent beams.
The number of beams per bin depends a bit on the UV-coverage which determines the naturally weighted beam area in pixels, typically about 50 pixels per beam.}. An odd number of spatial bins was specifically picked to allow a single bin at the map phase center as we 
wanted to be sure that strong source structure, often found at the phase center, was consistently sampled.  Testing with an $8\times8$ spatial grid gave nearly identical results, so having the phase center covered by four intersecting squares 
instead of just one may not be the disadvantage we initially suspected.

For the results discussed in this paper we track the entropy value after the 
signal-to-noise (SNR) ratio in the residual map drops below 6.0.  At higher SNR values, 
the dirty beam pattern in the residual map is strong enough that early rounds of CLEAN 
actually reduce the entropy until it begins to rise again below an SNR of about 6.0.  
The point at which the entropy begins to rise again can change with source structure and 
UV-coverage, but 6.0 works well for many simulated and real sources we tested.

\begin{figure*}[ht]
\centering
 \includegraphics[width=0.61\textwidth]{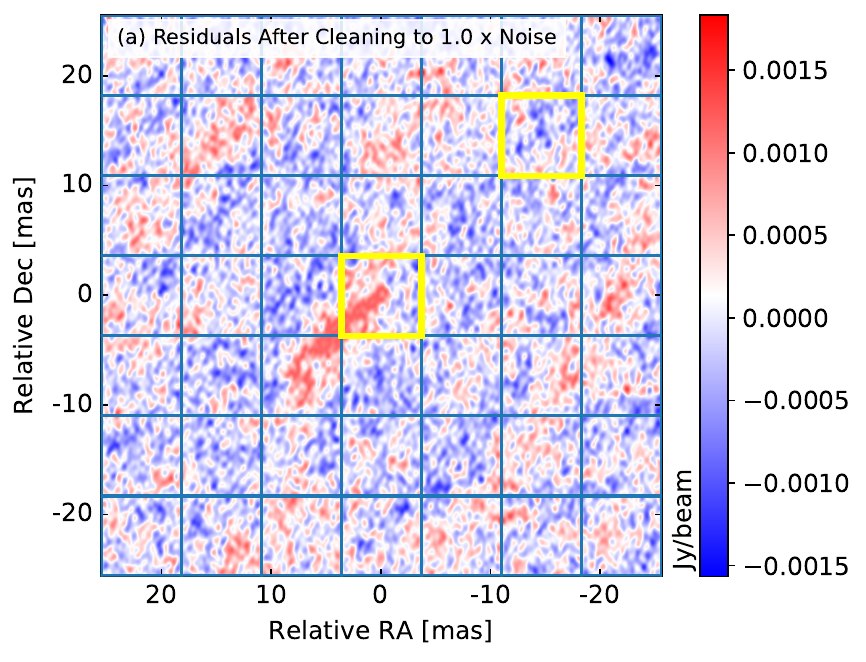}
 \includegraphics[width=0.34\textwidth]{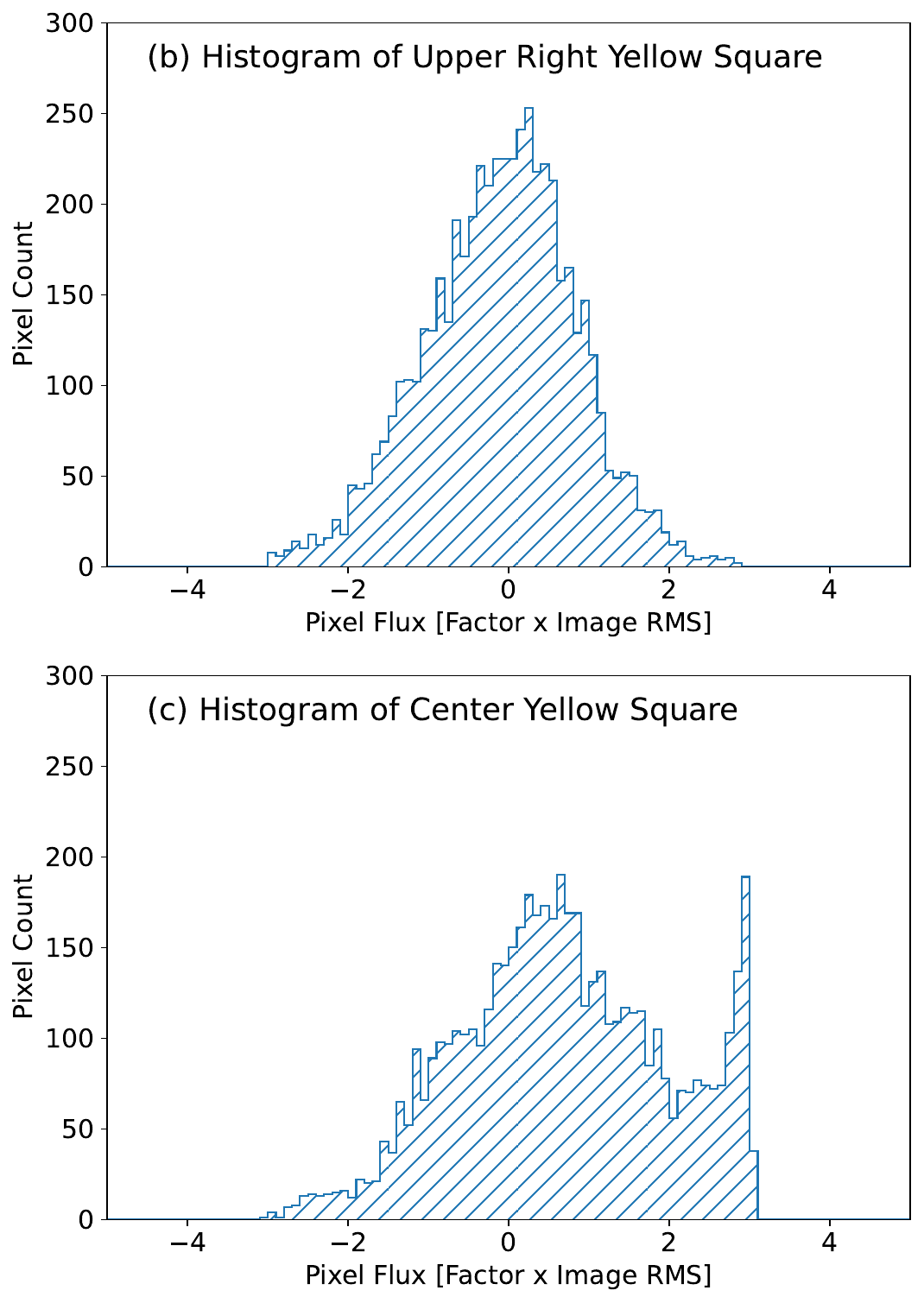}
 \includegraphics[width=0.475\textwidth]{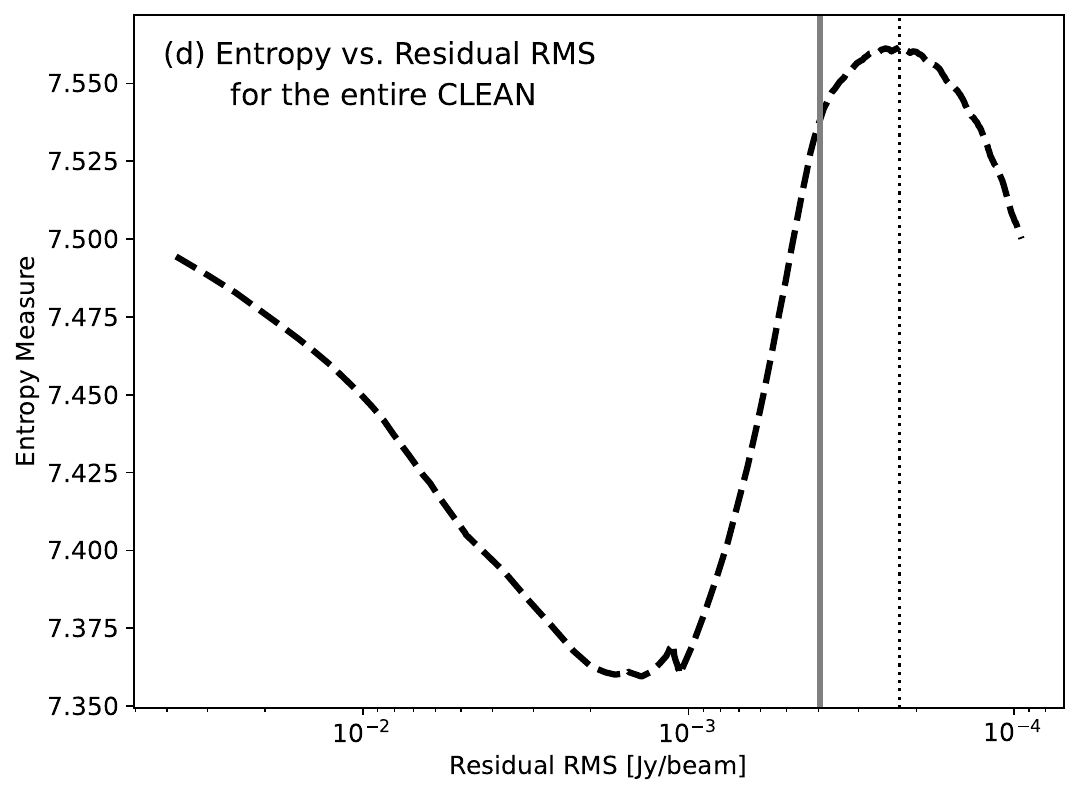}
 \includegraphics[width=0.505\textwidth]{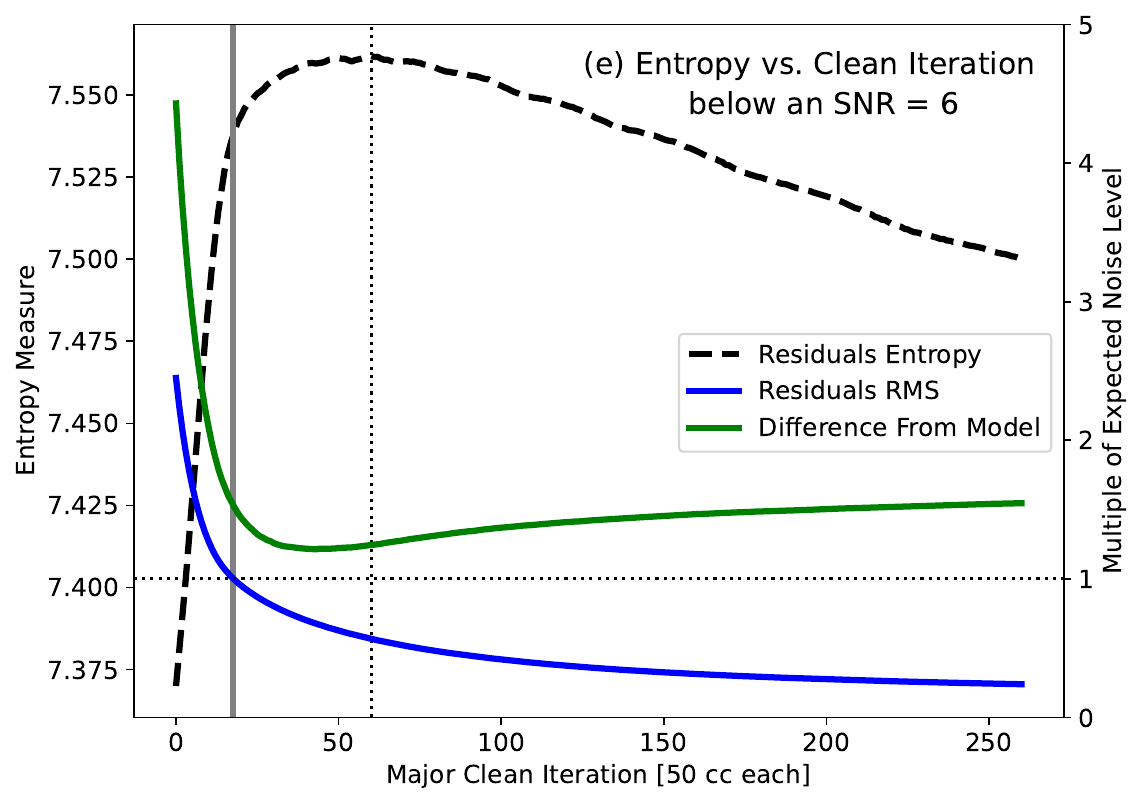}
\figcaption{\label{f:fig3}
  An illustration of our entropy calculation taken from the model shown in figure \ref{f:fig1} when the CLEAN has reached $1.0\times$ the known thermal noise.  Panel (a) is an expanded view of panel (d) from figure \ref{f:fig1} with our spatial entropy grid shown.  Two spatial boxes are highlighted with their flux histograms shown in panels (b) and (c). Panel (d) shows the entropy of the residuals vs. the residual RMS for the entire CLEAN.  The vertical gray and black dotted lines represent the expected noise level and the point of maximum entropy respectively.  Panel (e) shows in more detail how the entropy of the residuals, the residual RMS, and the on-source agreement between the current best CLEAN model and the known source model change in the final stages of CLEAN below an SNR of 6.0 ($\sim10^{-3}$ mJy/beam in panel (d)). Note that the right hand vertical axis of panel (e) applies to both the residual RMS (blue) and on-source agreement (green) lines. See the text for a more complete description.
}
\end{figure*}

Figure \ref{f:fig3} illustrates our entropy calculation by showing a close-up of
the residual map from figure \ref{f:fig1}d at the point where the RMS of the residuals
have reached $1.0\times$ the expected noise level in the simulation.  Note that significant
source structure remains, and further CLEAN iterations may be warranted.  Panels 
\ref{f:fig3}b and \ref{f:fig3}c show flux histograms of the two highlighted spatial bins 
in panel \ref{f:fig3}a.  Panel \ref{f:fig3}d shows how the entropy of the residual
map changes as a function of the RMS level of the residuals, which decrease from left
to right as the CLEAN proceeds.  The later iterations of CLEAN are shown in more
detail in panel \ref{f:fig3}e which illustrates the evolution of 
the entropy, residual RMS value, and agreement between the restored map and the known 
model with each CLEAN iteration if we were to stop at 
that point.  The image in panel \ref{f:fig3}a at $1.0\times$ the expected noise level 
corresponds to the gray vertical line
in panel \ref{f:fig3}e.  Cleaning beyond that point 
improves the agreement with the source model, but going far beyond that point causes the 
agreement to worsen.  We measure agreement between the restored CLEAN map at any given 
point and the known model by only computing the RMS difference on the pixels comprising 
known source structure, i.e. pixels either coincident with or adjacent to model component 
locations.  In this way, the green line in panel \ref{f:fig3}e only represents
on-source agreement and not agreement in other areas of the maps.

\subsection{Testing an Entropy Stopping criterion on Simulated Data}

\begin{figure*}[ht]
\centering
 \includegraphics[width=\textwidth]{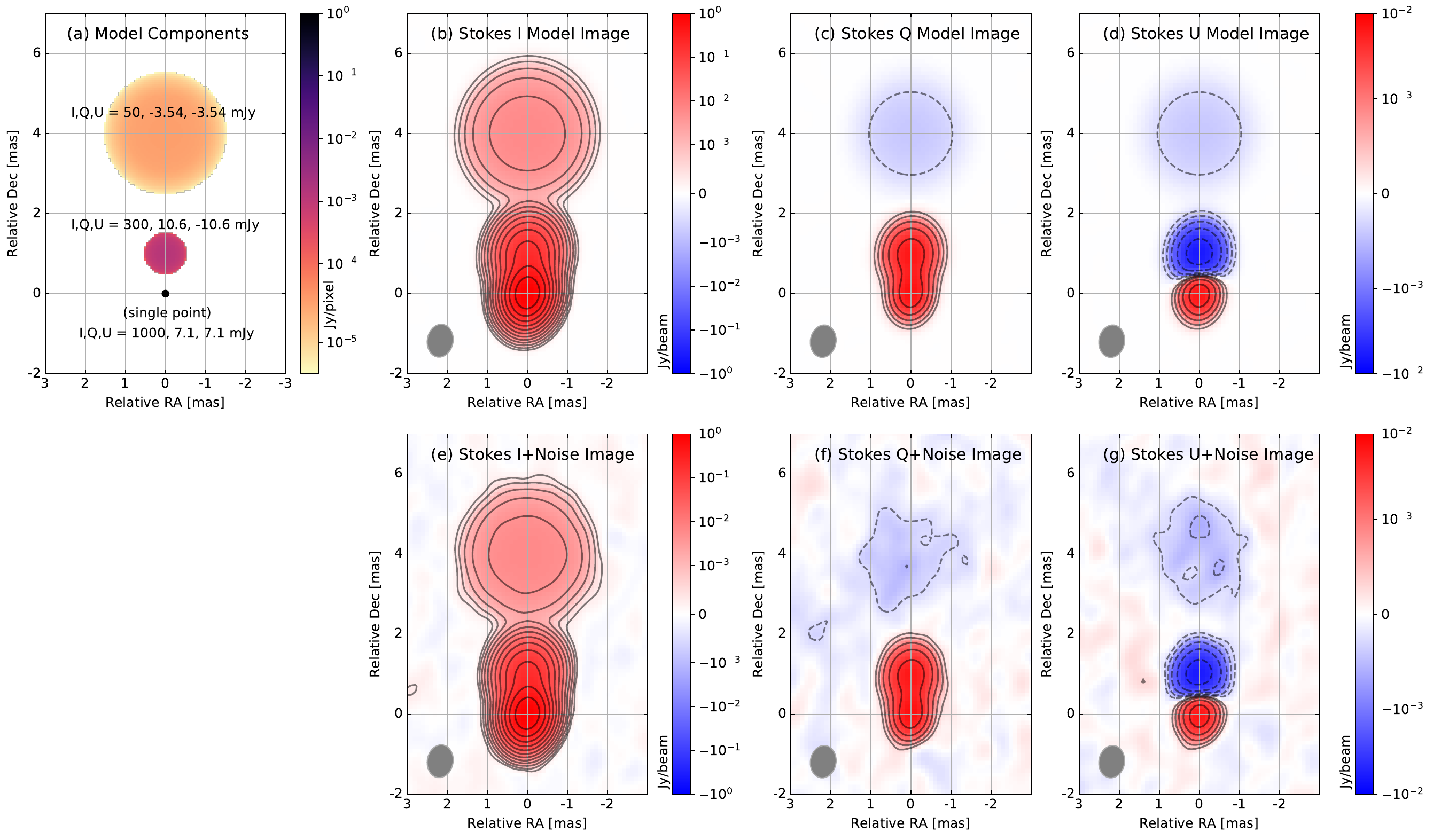}
\figcaption{\label{f:fig4}
  Illustration of our simulated jet model.  The model components are shown in panel (a) on a $0.05\times0.05$ mas$^2$ grid with the colorscale showing Stokes-$I$.  Total fluxes of the three major model features in $I$, $Q$, and $U$ are indicated on the plot.  Panels (b)-(d) show the model components convolved with our naturally weighted CLEAN beam in all three Stokes parameters, and Panels (e)-(g) show the same quantities but include a random realization of thermal noise in the simulated data.  All contours start at $\pm3.0\times$ the known thermal noise in this simulation of $0.091$ mJy/beam and increase in factor of $2.0$ steps.
}

\end{figure*}

To test entropy as a stopping criterion for CLEAN, we developed a simulated model
that mimics the source structure found in many core dominated, parsec-scale jets
with an extremely bright unresolved core, a nearby bright jet component that is 
partially resolved from the core, and a well separated, larger jet feature which
is significantly dimmer than the rest of the jet.  The two jet features have 
brightness profiles of homogeneous spheres and are constructed from point 
components on a grid, $0.05\times0.05$ milli-arcseconds [mas], which is $2\times$
finer than the mapping grid of $0.1\times 0.1$ mas that we use during imaging.  
For this reason
it is not possible, even in theory, for CLEAN to exactly reconstruct the
model source structure, providing a more realistic test of the imaging algorithm.
In addition to Stokes I, our model also had realistic linear polarization 
in the form Stokes $Q$ and $U$, as shown in figure~\ref{f:fig4}, and Stokes $V$ was 
set to zero.

Figure \ref{f:fig4}a shows the model components used to construct our 
simulated UV-data with the AIPS \citep{2003ASSL..285..109G} task UVMOD.  The model components replaced the existing data in three separate 15 GHz observations from the MOJAVE 
program spanning a range of UV-coverage: ``Poor'', ``Average'', and ``Good''.  The template file for our ``Poor'' UV-coverage was 3C\,273 (B1226$+$023) observed May 08, 2015 with a natural restoring beam of $1.33\times0.62$ mas$\times$mas at $-3.6$ degrees and estimated thermal noise of $0.115$ mJy/beam.  The template file for our ``Average'' UV-coverage was PKS 2209$+$236 observed June 17, 2017 with a restoring beam
of $1.04\times0.55$ mas$\times$mas at $-5.5$ degrees and an estimated thermal noise of $0.080$ mJy/beam.  The template file for our ``Good'' UV-coverage was 
3C\,380 (B1828$+$487) observed August 27, 2019 with a restoring beam of $0.83\times0.64$ mas$\times$mas at $-9.8$ degrees and an estimated thermal
noise of $0.091$ mJy/beam.  

\begin{figure*}[ht]
\centering
 \includegraphics[width=\textwidth]{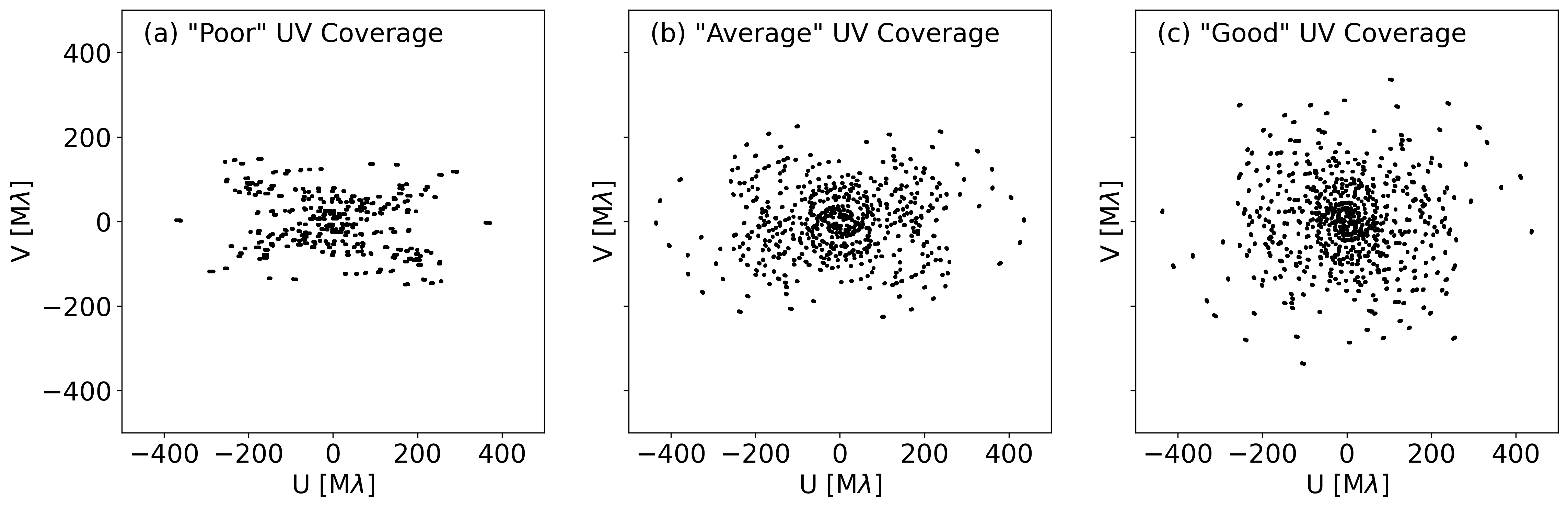} 
\figcaption{\label{f:fig5}
  Visual representation of the $UV$-coverage from each of the template UV data files described in the text. 
}
\end{figure*}

By substituting our simulated model components for the data found in real VLBA 
observations, we leverage the realistic UV-coverage and data weights found in
those observations and can more easily simulate expected noise levels and CLEAN
artifacts due to UV-coverage related issues. The three UV-coverages 
used are illustrated in figure \ref{f:fig5}, and for each of these UV-coverages
we constructed four different model types with the components shown in 
figure \ref{f:fig4}a oriented at $+45$ degrees, $0$ degrees, $-45$ degrees, and 
$-90$ degrees.  Finally, for each UV-coverage, orientation combination we 
created 10 random noise realizations with simulated noised added in the UV-plane calibrated to match the thermal noise from the
original VLBA observations that produced the UV-coverage.  As an illustration, figure \ref{f:fig4}, 
panels (b)-(d) show ideal model images with no significant noise\footnote{Actually, 
a low level of noise, $1\%$ of the noise levels in the full simulations, was 
introduced to avoid 
software issues we encountered with imaging zero noise data.}, and
panels (e)-(g) show images reconstructed from simulated data with noise added in
the UV-plane at a level matching the thermal noise in 
the original files.


For cleaning our simulations, we use a python script\footnote{Our python script can be downloaded
from the Zenodo service using the reference \citet{homan_daniel_2023_8416693}, and our simulated UV data can be
downloaded using the reference \citet{homan_daniel_2023_8416806}.} to drive Difmap \citep{2011ascl.soft03001S} 
and use the standard python astropy and scipy libraries \citep{2018AJ....156..123A,2022zndo....595738G} to load 
and analyze FITS files of the residual and restored maps at various stages throughout the 
CLEAN\footnote{In an early approach to this study \citep{2023AAS...24130141R}, we used a much slower 
CLEAN algorithm written by us 
entirely in python and obtained very similar results; however, we chose to work with 
Difmap for its much greater speed, reliability and wide-use by the 
VLBI community.}.  With this approach, we can collect statistics and 
monitor the progress of the CLEAN from start to finish.  This kind of scripting also 
allows us to use two strategies to improve the
early stages of CLEAN and thus minimize deconvolution errors early in the process.  First,
we perform the early rounds of the CLEAN, at very high SNR ($> 10$), with two levels of super-uniform weighting in Difmap to identify the locations of the brightest emission
as accurately as possible.  We then step down to regular uniform weighting
above an SNR of 6.0, and finally use natural weighting for the remainder of 
the CLEAN. These variable-weighting results are compared to a separate CLEAN which uses
straight natural weighting from the beginning, and the result with the lowest $\chi^2$
difference as computed from the naturally weighted UV-data at SNR = 6.0 is used from that point forward.  
Our second deconvolution improvement 
strategy is to repeatedly `re-CLEAN' our best result by reducing all clean components by 
just one percent (i.e. multiplying all clean component fluxes by 0.99).  This reduction
in clean component amplitude leaves new structure to be cleaned above SNR of 6.0, so we
repeat our earlier steps until the residuals reach an SNR of 6.0 again.  If the new
result is improved as measured by its $\chi^2$ agreement with the UV-data, we keep it, if not, 
we use the original CLEAN.

Both CLEAN improvement strategies seem to work well to improve early iterations of CLEAN
and avoid deconvolution errors that might otherwise be unrecoverable.  Because we are
always comparing the results of these strategies to more conservative approaches, there is no
danger that they will lead to a worse CLEAN, and in most cases, we find they improve the 
quality of the CLEAN overall.  The variable-weighting approach at high SNR has been used by 
the MOJAVE program for a long time, while the `re-CLEAN' strategy is something we 
discovered while working on this paper.  In the end, these strategies are about getting
the best possible deconvolution before we enter the final stages of CLEAN where a stopping
decision must be made.  This paper is fundamentally about the stopping criterion, so we do
not investigate general CLEAN improvement strategies here and save that for a future work.

\pagebreak

\section{Results and Discussion} \label{s:results}

\begin{figure*}[ht]
\centering
 \includegraphics[width=0.97\textwidth]{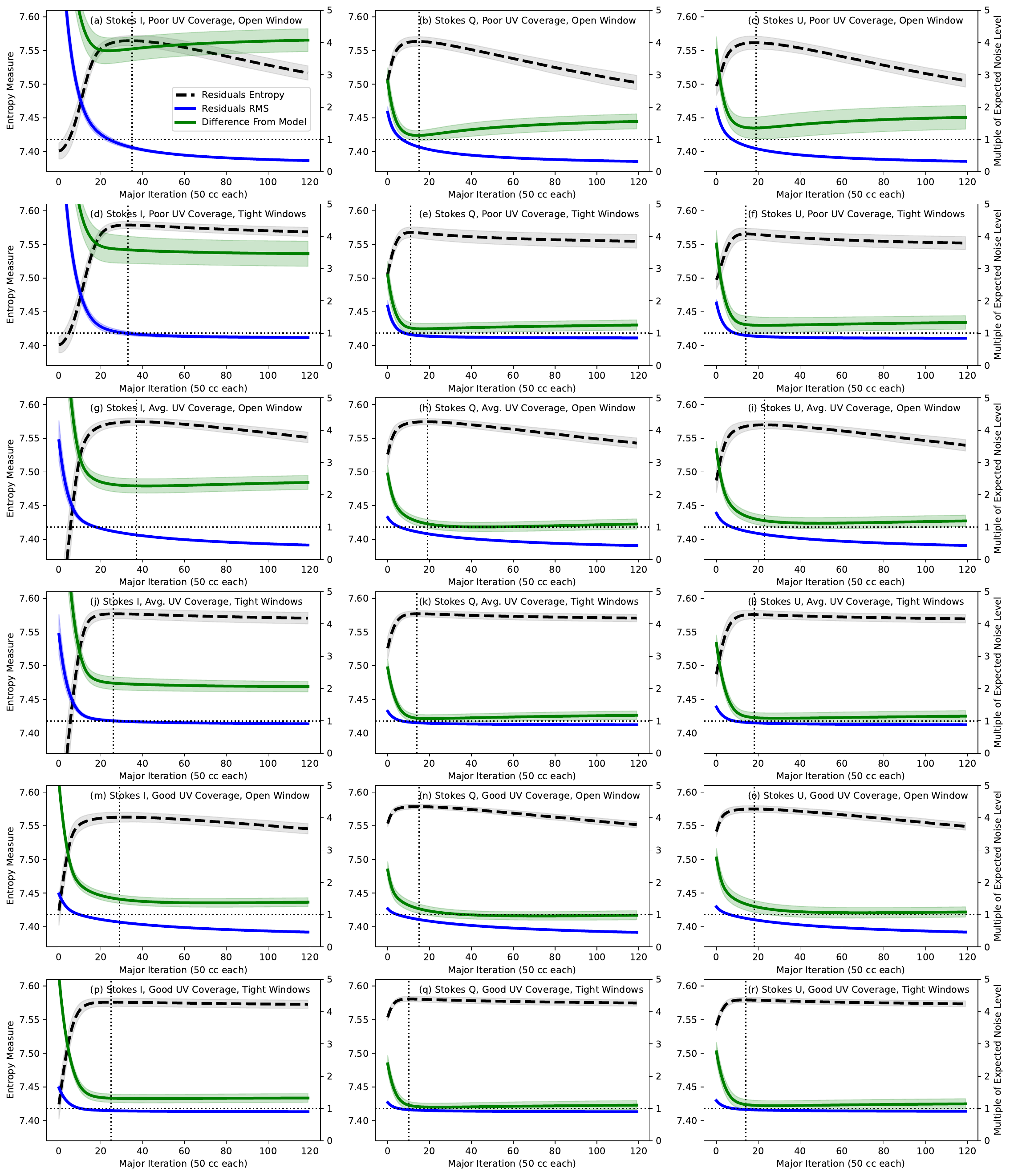}
\figcaption{\label{f:fig6}
  Plots of the change in entropy of the residuals, residual RMS, and on-source agreement between the restored CLEAN result and known model structure during the final rounds of CLEAN below SNR = 6.0.  Each plot represents the aggregate of 40 random simulations, as described in the text, with the central lines representing the mean quantities and the envelopes around those lines indicating the standard deviation.  Each pair of rows represents a different UV-coverage, with the first and second row showing CLEANs constrained either by a single large window or several windows tightly shaped to the known source structure.  Note that the right hand vertical axis of each panel applies to both the residual RMS (blue) and on-source agreement (green) lines.
}
\end{figure*}

Our results are shown in figure \ref{f:fig6} and each panel follows the same convention 
explained earlier for figure \ref{f:fig3}e.  Each row in figure \ref{f:fig6} shows
the CLEAN results for Stokes I, Q, and U from left to right.  The maximum entropy point
is marked with a vertical dashed line, and the horizontal dashed line indicates the 
known noise level in the simulated data.  Because we noticed only small differences 
between jet model orientations at a given UV-coverage, we have combined all $40$ random 
simulations for a given UV coverage into a single plot.  For example, the first line 
shows the aggregate of the 10 random simulations each at jet orientations of $+45$ degrees, $0$ degrees, $-45$ degrees, and $-90$ degrees.  The central lines for entropy 
(black, dashed line), RMS (blue line), and difference from model (green line) are the
average of all 40 models, with the envelopes indicating their standard deviation.

The reader will note that each of the three UV-coverages appears twice in figure \ref{f:fig6}.
The first row is for a CLEAN using a large, open cleaning window, covering approximately 50$\%$ 
of the mapping area as shown in figure \ref{f:fig1}.  The second row of plots is for
a CLEAN using `tight windows' drawn to surround the known jet structure with only about 
one beam of separation on each side.   We wanted to explore entropy as a potential 
stopping metric in both CLEAN scenarios, as they can have very different effects on 
large areas of the residual map where components are allowed in one case and not the other.

We find very consistent results across the board.  The maximum entropy point of the residuals produces a reasonable stopping point close to the optimum agreement between the final clean image and the model image on which the simulated data were based.  This is true for Stokes $I$, $Q$, and $U$ and for cleaning with a large single CLEAN window and with tight CLEAN windows.  We do notice that in the ``good'' UV-coverage case with a large single CLEAN window, the maximum entropy point on average lies a bit before the optimum, on-source agreement, but only by a small factor relative to the best achievable agreement with on-source structure.  

Figure \ref{f:fig6} shows the aggregate results of 40 simulations in each panel, and when we look through the individual simulations we do find cases where the peak entropy stops somewhat before or after the point of optimum, on-source agreement, but only six of the 360 cases represented in figure \ref{f:fig6} stood out to us as cases where CLEAN'ing more deeply would have produced meaningfully better on-source agreement.  These small number of cases were all in Stokes $Q$ and $U$ which show both positive and negative emission.  It is important to remember that the entropy calculated from the image residuals is subject to the same random statistical fluctuations as all noise, and we expect our entropy calculated at each step is only an estimate of the residual entropy one would find if you could repeat the experiment many times with different noise distributions.  

In real data, of course, one does not know the optimum source structure, and we only have an estimate of the thermal noise expected, as discussed earlier.  To make matters more interesting, additional effects, such as residual D-terms, gain errors, and deconvolution errors contribute to the noise budget.  In these cases, entropy of the residual map provides a useful statistic to monitor the progress of CLEAN in its final stages and determine the point at which further cleaning is likely to result in only recovering spurious structure and suppressing noise in the final image. We have applied these methods to many recent source observations from the MOJAVE program, and we find the maximum entropy stopping point for the residual map produces final clean images in good agreement with those produced by the current MOJAVE stopping criterion of one times the thermal noise estimate.  The maximum entropy stopping point typically occurs within about a factor of two of the thermal noise estimate, either above or below; however, in a small number of cases the maximum entropy stopping point comes considerably before the residuals reach the thermal noise estimate.  

In almost all of the small number of cases where the maximum entropy point comes well before the residuals reach the thermal noise estimate, there are good reasons to believe the maximum entropy stopping point is a more appropriate place to end the CLEAN.  These `early' stopping points seem to correspond to complex sources, with poor UV-coverage, like 3C\,273, or to cases where gain errors, residual D-terms or other polarization calibration issues contribute to random noise in the map.  Indeed, the application of this maximum entropy procedure during our testing has allowed us to uncover and fix a polarization issue in the data.  Even if entropy is not used as a stopping metric, these experiences suggest that entropy of the residuals is a useful diagnostic to track during deconvolution as it may reveal cases where data editing, calibration, and/or deconvolution issues exist and should be addressed.

In rare cases, UV-coverage and source structure combine to create regular patterns in the residual image, often mirroring the source structure itself when the CLEAN proceeds with only the constraint of a single, large cleaning window.  In these cases, the residual entropy may reach a maximum when additional cleaning may still be useful in recovering source structure.  We have found some advantage to changing the gridding size of the entropy calculation to avoid coincidence with repeating structure in the residual map and/or simply cleaning the map with a larger image size.  An alternative is to use tight cleaning windows, if one has good a-priori knowledge of the source structure, which usually avoids this issue.

\section{Summary} \label{s:summary}

We propose measuring the entropy of the residual map, calculated using both the spatial locations and fluxes of the pixels, as a diagnostic and potential stopping metric for CLEAN algorithm.  We've applied this technique to simulated VLBA data where we know the source structure and true noise distribution, and we find that tracking the rise of the entropy of the residuals during final rounds of CLEAN provides a natural stopping point at maximum entropy that, in most cases, recovers the source structure near or at the point of optimum agreement between the final CLEAN map and known source model. This approach seems to work well for both total intensity, Stokes $I$, and linear polarization, Stokes $Q$ and $U$.

In testing this approach on real MOJAVE data, where we do not know the true source structure and can only estimate the thermal noise in our data, the maximum entropy stopping point typically lies within a factor of two above or below the estimated thermal noise. In most of the small number of cases where the maximum entropy point was reached well in advance of the estimated thermal noise, real issues with deconvolution or data calibration were apparent, and it was unlikely we could have recovered meaningful source structure by cleaning further.  In this way, monitoring entropy of the residuals during deconvolution can serve not only as a stopping metric but also as a diagnostic, potentially revealing problems with the imaging or data calibration. 

\begin{acknowledgments}
JSR was supported by the Reid and Polly Anderson Endowment of Denison University.
This research has made use of data from the MOJAVE database that is maintained by the MOJAVE team \citep{2018ApJS..234...12L}.  
%
This research has made use of NASA's Astrophysics Data System.
\end{acknowledgments}


\bibliography{EntropyCLEAN}{}
\bibliographystyle{aasjournal}



\end{document}